\documentclass[12pt]{article}
\usepackage[dvips]{epsfig}
\usepackage{amsfonts,amsmath,amsbsy}
\def\bbox#1{\boldsymbol#1}
\def\upA{\uparrow}
\def\dnA{\downarrow}
\def\B{{\cal B}}

\def\P{{\cal P}}

\def\a{{\cal A}}

\def\DSAS{\Delta_{\text{SAS}}}
\def\dx{d^{2}x}
\def\dy{d^{2}y}

\def\bpmatrix{\left(\begin{array}{c}}
\def\epmatrix{\end{array}\right)}
\def\bvector{\left(\begin{array}{c}}
\def\evector{\end{array}\right)}
\def\eff{\text{eff}}

\def\vor{\text{vor}}
\def\sky{\text{sky}}
\def\spin{\text{spin}}
\def\ppin{\text{ppin}}

\def\CB{composite-boson\ }
\def\CBs{composite bosons}
\def\LLL{lowest Landau level}
\def\F{{\mathfrak S}}
\def\Laugh{\F_{\text{LN}}[\bbox{x}]}
\begin{document}
\title{Improved Composite-Boson Theory \\
of\\
Monolayer and Bilayer Quantum Hall Ferromagnets
}
\author{Z. F. Ezawa\\
Department of Physics, Tohoku University, Sendai 980, Japan
}
\maketitle
\begin{abstract}
An improved composite-boson theory of quantum Hall ferromagnets is formulated 
both for the monolayer and bilayer systems.  In this scheme the field operator 
describes solely the physical degrees of freedom representing the deviation 
from the ground state.  Skyrmions are charged excitations confined to the 
\LLL.  By evaluating the excitation energy of one skyrmion in the 
interlayer-coherent phase it is shown that the bilayer QH state becomes 
stabler as the interlayer density difference becomes larger.
\newline
\newline
{\sl PACS:}\ 73.40.Hm, 73.20.Dx, 73.40.-c, 75.10.-b\newline
{\sl Keywords:}\ quantum Hall effect, quantum coherence, bilayer electron 
system, skyrmions
\end{abstract}
\baselineskip17pt

\section{Introduction}

The quantum Hall (QH) effect is a remarkable macroscopic quantum 
phenomenon in the two-dimensional electron system \cite{FQHEbook}.  The underlying 
physics is understood by the composite-boson (CB) picture 
\cite{LGCSx,GirvinAA,ReadA,RajaramanCB} or by the composite-fermion picture 
\cite{JainCF,OtherCF}.  There are several new approaches \cite{EzaICBa,NewApproCP} to 
the problem by extending these pictures.

When the spin degree of freedom is taken into account, a spin coherence 
develops spontaneously and turns the QH system into a QH ferromagnet, provided 
the Zeeman effect is reasonablly small.  Quasiparticles are skyrmions 
\cite{SkyrmQH}, whose existence has been established experimentally \cite{SkyrmExper}.  
On the other hand, a pseudospin (interlayer) coherence develops spontaneously 
in certain bilayer QH systems \cite{EIcoher,EzaIQC,MG}.  Some of its 
characteristic behaviors have already been observed experimentally 
\cite{Sheena,SawaPRLa}.  In a coherent state the phase and number differences are 
observables simultaneously.  The phase difference is controlled by applying 
the parallel magnetic field \cite{Sheena} while the number difference is 
controlled by applying bias voltages to the bilayer system \cite{SawaPRLa}.  

In this paper we analyze skyrmion excitations in monolayer and bilayer 
QH ferromagnets.  We use the improved \CB\ (CB) theory \cite{EzaICBa}, proposed 
based on a suggestion due to Girvin \cite{GirvinAA}, Read \cite{ReadA} and Rajaraman 
et al. \cite{RajaramanCB}. The advantages of the improved scheme are that the 
field operator describes solely the physical degrees of freedom representing 
the deviation from the ground state and that the semiclassical property of 
excitations is determined directly by the microscopic wave function.  

In bilayer QH systems we have two types of skyrmions; the 
\textit{spin-skyrmion} associated with the spin coherence and the 
\textit{pseudospin-skyrmion} associated with the pseudospin coherence.  One 
skyrmion consists of a pair of charged excitations on the two layers, 
and hence acquires a capacitive charging energy, whichever type of skyrmion it 
may be.  We have predicted \cite{EIcoher,EzaIQC} that the interlayer-coherent 
state continues to exist even if the electron densities are made arbitrarily 
unbalanced between the two quantum wells, as has been confirmed experimentally 
\cite{SawaPRLa}.  Furthermore it has been revealed \cite{SawaPRLa} that the activation 
energy increases as the density imbalance increases.  We explain this 
characteristic behavior by evaluating the excitation energy of one skyrmion as 
a function of the density imbalance:  The major energy is the capacitive 
charging energy we mentioned.  Throughout the paper we use the natural units 
$\hbar =c=1$.

\section{Bosonization}

We summarize the idea of the improved CB theory applied to the 
monolayer spin-frozen QH state.  We denote the electron field by $\psi (\bbox{x})$ and 
its position by the complex coordinate, which we normalize as $z=(x+iy)/2\ell _{B}$ 
with  $\ell _{B}$ the magnetic length.  Any state $|\F\rangle $ at the filling factor 
$\nu =1/m$ ($m$ odd) is represented by the wave function,
\begin{equation}
\F[\bbox{x}] \equiv  \langle 0|\psi (\bbox{x}_{1})\cdots \psi (\bbox{x}_{N})|\F\rangle  = \omega [z]\Laugh ,
\label{WaveElect}
\end{equation}
where $\Laugh$ is the Laughlin function,
\begin{equation}
\Laugh=\prod _{r<s}(z_{r}-z_{s})^{m}\exp[-\sum _{r=1}^{N}|z_{r}|^{2}] ,
\end{equation}
and $\omega [z]\equiv \omega (z_{1},z_{2},\cdots ,z_{N})$ is an analytic function symmetric in all $N$ 
variables.  The mapping from the fermionic wave function $\F[\bbox{x}]$ to the 
bosonic function $\omega [z]$ defines a bosonization.  We call the underlying boson 
the \textit{dressed composite boson} and denote its field operator by $\varphi (\bbox{x})$.  
The field operator turns out to be the one considered first by Read \cite{ReadA} 
and revived recently by Rajaraman et al. \cite{RajaramanCB}.  It follows that
\begin{equation}
\F_{\varphi }[\bbox{x}] \equiv  \langle 0|\varphi (\bbox{x}_{1})\cdots \varphi (\bbox{x}_{N})|\F\rangle  = \omega [z] .
\label{WaveFunctDress}
\end{equation}
The Laughlin state is represented by $\F_{\varphi }[\bbox{x}]=1$.  A typical quasiparticle 
(vortex) state \cite{LaughlinA} is described by $\F_{\varphi }[\bbox{x}]=\prod _{r}^{N}z_{r}$, leading to 
$\langle \varphi (\bbox{x})\rangle =z$ in the semiclassical approximation.  This is a highly nontrivial 
constraint, which turns out to determine all semiclassical properties of the 
vortex excitation.  Consequently, the field operator $\varphi (\bbox{x})$ describes solely 
the physical degrees of freedom representing the deviation from the ground 
state, and the semiclassical property of excitations is determined directly by 
the microscopic wave function (\ref{WaveElect}).

We now construct the improved CB theory explicitly.  We start with the 
kinetic Hamiltonian for planar electrons in external magnetic field 
$(0,0,-B)$,
\begin{equation}
H_{K}= {1\over 2M}\int \dx \psi ^{\dagger }(\bbox{x})(P_{x}-iP_{y})(P_{x}+iP_{y})\psi (\bbox{x}) ,
\label{HamilKinem}
\end{equation}
where $P_{j}=-i\partial _{j}+eA_{j}^{\text{ext}}$ is the covariant momentum with 
$A_{j}^{\text{ext}}={1\over 2}\varepsilon _{jk}x_{k}B$.  Here, $\varepsilon _{12}=-\varepsilon _{21}=1$ and $\varepsilon _{11}=\varepsilon _{22}=0$.  We denote 
the electron density by $\rho (\bbox{x})=\psi ^{\dagger }(\bbox{x})\psi (\bbox{x})$.  The Coulomb term is
\begin{equation}
H_{C} = {1\over 2}\int \dx\dx V(\bbox{x}-\bbox{y})\varrho  (\bbox{x})\varrho  (\bbox{y}) ,
\label{HamilCoulo}
\end{equation}
with $V(\bbox{x})=(e^{2}/\varepsilon )|\bbox{x}|^{-1}$, where $\varrho  (\bbox{x})\equiv \rho (\bbox{x})-\rho _{0}$ is the density deviation from 
its average $\rho _{0}\equiv \langle \rho (\bbox{x})\rangle $.  It is normalized to vanish, $\langle H_{C}\rangle =0$, on the 
homogeneous ground state.

At sufficiently low temperature all relevant excitations are those 
confined to the \LLL.  The condition is that the kinetic energy (\ref{HamilKinem}) 
is quenched on the state,
\begin{equation}
(P_{x}+iP_{y})\psi (\bbox{x})|\F\rangle =0 .
\label{LLLcondiElect}
\end{equation}
We call it the lowest-Landau-level (LLL) condition.

We define the \textit{bare} CB field by way of an operator phase 
transformation of the electron field $\psi (\bbox{x})$, 
\begin{equation}
\phi (\bbox{x})=e^{-i\Theta (\bbox{x})}\psi (\bbox{x}) .
\label{BareField}
\end{equation}
The phase field $\Theta (\bbox{x})$ is chosen to attach $m$ units of Dirac flux quanta 
$2\pi /e$ to each electron via the relation,
\begin{equation}
\varepsilon _{ij}\partial _{i}\partial _{j}\Theta (\bbox{x})=2\pi m\rho (\bbox{x}). 
\label{PhaseField}
\end{equation}
When $m$ is odd, $\phi (\bbox{x})$ is proved to be a bosonic operator.  The bare 
composite boson is the one familiar in literatures 
\cite{LGCSx,EIcoher,EzaIQC}.

We proceed to define the {\it dressed} CB field $\varphi (\bbox{x})$,
\begin{equation}
\varphi (\bbox{x})=e^{-\a(\bbox{x})}\phi (\bbox{x}) .
\label{DressField}
\end{equation}
Here, the hermitian field $\a(\bbox{x})$ is to be determined so that the basic 
formula (\ref{WaveFunctDress}) is obtained for the wave function.  Substituting 
(\ref{DressField}) and (\ref{BareField}) into (\ref{HamilKinem}), the kinetic Hamiltonian is 
transformed into 
\begin{equation}
H_{K} = {1\over 2M}\int d^{2}x \varphi ^{\ddag }(\bbox{x})(\P_{x}-i\P_{y})(\P_{x}+i\P_{y})\varphi (\bbox{x}) ,
\label{HamilCB}
\end{equation}
where $\varphi ^{\ddag }(\bbox{x})\equiv \varphi ^{\dagger }(\bbox{x})e^{2\a(\bbox{x})}$, with which $\rho (\bbox{x})=\psi ^{\dagger }(\bbox{x})\psi (\bbox{x})=\varphi ^{\ddag }(\bbox{x})\varphi (\bbox{x})$.  The 
covariant momentum is
\begin{equation}
\P_{j} =  -i\partial _{j} + eA_{j}^{\text{ext}}(\bbox{x}) + \partial _{j}\Theta (\bbox{x}) - i\partial _{j}\a(\bbox{x}) .
\label{CovarMomenRx}
\end{equation}
By a judicious choice of $\a(\bbox{x})$ we are able to bring the LLL condition 
(\ref{LLLcondiElect}) into a simple form,
\begin{equation}
(\P_{x}+i\P_{y})\varphi (\bbox{x})|\F\rangle =-{i\over \ell _{B}}{\partial \over \partial z^{*}}\varphi (\bbox{x})|\F\rangle  = 0 .
\label{LLLcondiDress}
\end{equation}
Indeed, when we choose $\partial _{i}\a(\bbox{x})=\varepsilon _{ij}[eA_{j}^{\text{ext}}(\bbox{x})+\partial _{j}\Theta (\bbox{x})]$, the covariant 
momentum reads
\begin{equation}
\P_{j} =  -i\partial _{j} - (\varepsilon _{jk} + i\delta _{jk})\partial _{k}\a(\bbox{x}),
\label{CovarMomenR}
\end{equation}
from which the LLL condition (\ref{LLLcondiDress}) follows trivially.  By using 
(\ref{PhaseField}), the above definition of $\a(\bbox{x})$ leads to
\begin{equation}
\bbox{\nabla }^{2}\a(\bbox{x}) = 2\pi m \biggl(\rho (\bbox{x})-{eB\over 2\pi m}\biggr) .
\label{SpinA}
\end{equation}
It is easy to see that bare composite bosons feel the effective magnetic field 
$\B_{\text{eff}}=e^{-1}\bbox{\nabla }^{2}\a(\bbox{x})$ and that $\B_{\eff}$ vanishes on the homogeneous ground 
state where $\langle \rho (\bbox{x})\rangle =\rho _{0}$.  It follows that the homogeneous ground state is 
realized only at the filling factor $\nu \equiv 2\pi \rho _{0}/eB=1/m$, where \CBs\ undergo bose 
condensation.  Eq.(\ref{SpinA}) is solved as
\begin{equation}
\a(\bbox{x}) = m\int \dy \ln\biggl({|\bbox{x}-\bbox{y}|\over 2\ell _{B}}\biggr) \varrho  (\bbox{y}) ,
\label{SpinB}
\end{equation}
in terms of the density deviation $\varrho  (\bbox{x})$ at $\nu =1/m$.  It is interpreted that 
the new CB field (\ref{DressField}) is obtained by dressing the bare field $\phi (\bbox{x})$ 
with a cloud of the effective magnetic field generated by $\a(\bbox{x})$, and hence 
we have called it the dressed field.  

Solving the LLL condition (\ref{LLLcondiDress}) we find that the $N$-body 
wave function $\F_{\varphi }[\bbox{x}]$ is an analytic function as in (\ref{WaveFunctDress}).  It is 
an easy exercise to derive the relation \cite{RajaramanCB},
\begin{align*}
\varphi ^{\ddag }(\bbox{x}_{1})\cdots \varphi ^{\ddag }(\bbox{x}_{N})|0\rangle  = \Laugh \psi ^{\dagger }(\bbox{x}_{1})\cdots \psi ^{\dagger }(\bbox{x}_{N})|0\rangle  .
\end{align*}
Because of this relation the function $\omega [z]$ in the wave function (\ref{WaveElect}) 
is given precisely by the formula (\ref{WaveFunctDress}).

One might question the hermiticity of the theory \cite{RajaramanCB}, since 
the covariant momentum (\ref{CovarMomenR}) has an unusual expression.  Analyzing 
the Lagrangian density we find that the canonical conjugate of $\varphi (\bbox{x})$ is not 
$i\varphi ^{\dagger }(\bbox{x})$ but $i\varphi ^{\ddag }(\bbox{x})\equiv i\varphi ^{\dagger }(\bbox{x})e^{2\a(\bbox{x})}$.  It implies that the hermiticity is 
defined together with the measure $e^{2\a(\bbox{x})}$.  Such a measure has arisen since 
the transformation (\ref{DressField}) is not unitary.  The covariant momentum 
(\ref{CovarMomenR}) is hermitian together with this measure.  It is instructive to 
rewrite the kinetic Hamiltonian (\ref{HamilCB}) as
\begin{equation}
H_{K} = {\omega _{c}\over 2}\int \dx \biggl({\partial \over \partial z^{*}}\varphi (\bbox{x})\biggr)^{\dagger }e^{2\a(\bbox{x})}{\partial \over \partial z^{*}}\varphi (\bbox{x}) ,
\label{HamilMono}
\end{equation}
which is manifestly hermitian.  

\section{Monolayer QH Ferromagnets}

We analyze the QH system with the spin degree of freedom.  The electron 
field $\psi ^{\alpha }(\bbox{x})$ is labeled by the spin index $\alpha =\upA ,\dnA $.  Bare and dressed CB 
fields $\phi ^{\alpha }(\bbox{x})$ and $\varphi ^{\alpha }(\bbox{x})$ are defined by (\ref{BareField}) and (\ref{DressField}), 
where $\Theta (\bbox{x})$ and $\a(\bbox{x})$ are defined by (\ref{PhaseField}) and (\ref{SpinB}) with 
$\rho (\bbox{x})=\Psi ^{\dagger }(\bbox{x})\Psi (\bbox{x})=\Phi ^{\dagger }(\bbox{x})e^{2\a(\bbox{x})}\Phi (\bbox{x})$.  Here, $\Psi (\bbox{x})$ and $\Phi (\bbox{x})$ are two-component 
fields made of $\psi ^{\alpha }(\bbox{x})$ and $\varphi ^{\alpha }(\bbox{x})$.  We decompose the bare CB field into the 
U(1) field $\phi (\bbox{x})$ and the SU(2) field $n^{\alpha }(\bbox{x})$, $\phi ^{\alpha }(\bbox{x})=\phi (\bbox{x})n^{\alpha }(\bbox{x})$, by requiring 
$\sum _{\alpha }n^{\alpha \dagger }n^{\alpha }=1$.  Thus,
\begin{equation}
\varphi ^{\alpha }(\bbox{x}) = e^{-\a(\bbox{x})}\phi ^{\alpha }(\bbox{x})=e^{-\a(\bbox{x})}\phi (\bbox{x})n^{\alpha }(\bbox{x}) .
\label{BasicFormu}
\end{equation}
The field $n^{\alpha }(\bbox{x})$ is the complex-projective field \cite{Dadda}, whose overall 
phase has been removed and given to the U(1) field $\phi (\bbox{x})$.  

The kinetic Hamiltonian reads
\begin{equation}
H_{K} = {1\over 2M}\int d^{2}x \Phi ^{\ddag }(\bbox{x})(\P_{x}-i\P_{y})(\P_{x}+i\P_{y})\Phi (\bbox{x}) .
\label{HamilBL}
\end{equation}
The LLL condition is
\begin{equation}
(\P_{x}+i\P_{y})\Phi (\bbox{x})|\F\rangle =-{i\over \ell _{B}}{\partial \over \partial z^{*}}\Phi (\bbox{x})|\F\rangle  = 0 .
\label{LLLcondiBL}
\end{equation}
Therefore the CB wave function $\F_{\varphi }[z]$ is symmetric and analytic in $N$ 
coordinates $z_{i}$, with which the electron wave function is given by 
$\F[\bbox{x}]=\F_{\varphi }[z]\Laugh$.  

We use the index $a=x,y,z$ for the spin component.  The spin density is
\begin{equation}
S^{a}(\bbox{x})={1\over 2}\Psi ^{\dagger }(\bbox{x})\tau ^{a}\Psi (\bbox{x})={1\over 2}\rho (\bbox{x})\bbox{s}^{a}(\bbox{x}), 
\label{SpinDensi}
\end{equation}
where $\tau ^{a}$ is the Pauli matrix and $\bbox{s}^{a}(\bbox{x})$ is the normalized spin field, or 
the nonlinear sigma field,
\begin{equation}
s^{a}(\bbox{x})=\bbox{n}^{\dagger }(\bbox{x})\tau ^{a}\bbox{n}(\bbox{x}), 
\label{SigmaField}
\end{equation}
with $\bbox{n}(\bbox{x})$ the two-component field made of $n^{\upA }(\bbox{x})$ and $n^{\dnA }(\bbox{x})$.  The Zeeman 
term is
\begin{equation}
H_{Z} = -{1\over 2}g^{*}\mu _{B}B \int d^{2}x \rho (\bbox{x})s^{z}(\bbox{x}), 
\end{equation}
where $g^{*}$ is the gyromagnetic factor and $\mu _{B}$ the Bohr magneton.  Each Landau 
level contains two energy levels for spin-up and spin-down states with the 
one-particle gap energy $g^{*}\mu _{B}B$.  The ground state $|g_{0}\rangle $ is unique, whose 
wave function is
\begin{equation}
\F^{\spin}_{\text{g}}[\bbox{x}] = \prod _{r}\bpmatrix 1\\ 0\epmatrix_{r}^{\spin} \Laugh,
\label{GrounWave}
\end{equation}
where the two-component spinor is common to all electrons, representing the 
spin-up polarization.

The effective Hamiltonian governing perturbative fluctuations of the 
sigma field has been derived previously \cite{EzaIQC,MG},
\begin{equation}
H_{\text{eff}}= {1\over 2}\rho _{s}\int \dx\biggl\{[\partial _{k}\bbox{s}(\bbox{x})]^{2} - {\rho _{0}\over \rho _{s}}g^{*}\mu _{B}B s^{z}(\bbox{x})\biggr\}.
\label{EnergChangSPN}
\end{equation}
The first term represents the spin stiffness with $\rho _{s}=\nu e^{2}/(16\sqrt {2\pi }\varepsilon \ell _{B})$.  
Perturbative excitations are charge neutral.  We consider the vanishing limit 
of the Zeeman term ($g^{*}=0$).  In this case the ground state is given by an 
arbitrary constant sigma field, $\bbox{s}(\bbox{x})$=constant.  All spins are polarized into 
one arbitrary direction.  There exists a degeneracy in the ground states.  The 
choice of a ground state implies a spontaneous magnetization, or a QH 
ferromagnetism.  When a continuous symmetry is spontaneously broken, there 
arises a gapless mode known as the Goldstone mode.  Quantum coherence develops 
spontaneously.  Actually, due to the Zeeman term all spins are polarized into 
the $z$ axis.  As far as the Zeeman effect is small enough, the system is 
still considered as a QH ferromagnet with a finite coherent length.

\section{Topological Excitations}

We analyze topological excitations on the QH ferromagnet.  The 
semiclassical analysis is powerful when the CB wave function $\F_{\varphi }[z]$ is 
factorizable,
\begin{equation}
\F^{\spin}[\bbox{x}] = \prod _{r}\bpmatrix \omega ^{\upA }(z_{r})\\ \omega ^{\dnA }(z_{r})\epmatrix_{\kern-1pt r}^{\spin} \Laugh.
\label{SkyrmWave}
\end{equation}
In this case the one-point function is given by $\langle \varphi ^{\alpha }(\bbox{x})\rangle \equiv \omega ^{\alpha }(z)$.  Based on the 
formula (\ref{BasicFormu}) it is parametrized as
\begin{equation}
e^{-\a(\bbox{x})}e^{i\chi (\bbox{x})} \sqrt {\rho _{0}+\varrho  (\bbox{x})}n^{\alpha }(\bbox{x}) = \omega ^{\alpha }(z) ,
\label{ClassGenerB}
\end{equation}
since $|\phi (\bbox{x})|^{2}=\rho _{0}+\varrho  (\bbox{x})$.  Here and hereafter all fields are classical fields.  
The Cauchy-Riemann equation for (\ref{ClassGenerB}) yields \cite{EzaIQC},
\begin{equation}
\varrho  (\bbox{x}) = -\nu Q_{0}(\bbox{x}) + {\nu \over 4\pi }\bbox{\nabla }^{2}\ln\biggl(1+{\varrho  (\bbox{x})\over \rho _{0}}\biggr) ,
\label{SolitEquat}
\end{equation}
where use was made of (\ref{SpinA}), and $Q_{0}(\bbox{x})$ is the topological charge density 
whose explicit form is given later.  A nontrivial density deviation $\varrho  (\bbox{x})$ 
is induced around a topological soliton according to this equation, which we 
name the soliton equation.  Since the soliton equation is a direct consequence 
of the semiclassical LLL condition (\ref{ClassGenerB}), it is interpreted that the 
density deviation occurs in order to confine topological excitations within 
the \LLL.  

The excitation carries a quantized electric charge,
\begin{equation}
q_{e}=-e\int \dx \varrho  (\bbox{x}) = e\nu \int \dx Q_{0}(\bbox{x}) = \nu qe ,
\label{TotalCharg}
\end{equation}
as follows from (\ref{SolitEquat}), where $q=\int \dx Q_{0}(\bbox{x})$ is the topological charge.  
We explicitly consider solitons with $q=1$.  One soliton carries a fractional 
charge, say, $q_{e}={1\over 3}e$ at $\nu =1/3$, as has been experimentally observed 
\cite{AnyonExperA}.  We now show that there are topological solitons, vortices and 
skyrmions, associated with the U(1) and SU(2) parts of the CB field $\Phi (\bbox{x})$.  

\subsection{Vortex Excitations}

The U(1) excitation is generated on the ground state when 
$\partial _{k}\chi (\bbox{x})\not=0$ and $n^{\alpha }(\bbox{x})=\text{constant}$ in (\ref{ClassGenerB}).  The 
complex-projective field is trivial, $n^{\upA }=1$ and $n^{\dnA }=0$, representing a spin-up 
polarized ground state (\ref{GrounWave}).  It follows from (\ref{ClassGenerB}) that the 
topological charge density is
\begin{equation}
Q_{0}(\bbox{x})\equiv Q^{V}_{0}(\bbox{x})={1\over 2\pi }\varepsilon _{jk}\partial _{j}\partial _{k}\chi (\bbox{x}), 
\end{equation}
as represents the vorticity.  

The simplest excitation is given by one vortex sitting at $\bbox{x}=0$, whose 
wave function is (\ref{SkyrmWave}) with $\omega ^{\upA }(z)=\sqrt {\rho _{0}}z$ and $\omega ^{\upA }(z)=0$.  The 
topological charge is concentrated at the vortex center, $Q^{V}(\bbox{x})=\delta (\bbox{x})$.  An 
approximate solution of the soliton equation (\ref{SolitEquat}) reads
\begin{equation}
\varrho  _{\vor}(\bbox{x}) \simeq  - \rho _{0}\biggl(1+{\sqrt {2}r\over \ell _{B}}-{r^{2}\over 3\ell _{B}^{2}}\biggr)e^{-\sqrt {2}r/\ell _{B}} .
\label{BetteAppro}
\end{equation}
A vortex is a hole made in the condensate of composite bosons.

\subsection{Skyrmion Excitations}

The SU(2) excitation is generated on the ground state when 
$\chi (\bbox{x})=\text{constant}$ and $\partial _{k}n^{\alpha }(\bbox{x})\not=0$ in (\ref{ClassGenerB}).  The 
complex-projective field is solved as
\begin{equation}
n^{\alpha }(\bbox{x}) = {\omega ^{\alpha }(z)\over \sqrt {|\omega ^{\upA }(z)|^{2}+|\omega ^{\dnA }(z)|^{2}}} ,
\label{GenerSkyrm}
\end{equation}
with arbitrary analytic functions $\omega ^{\alpha }(z)$.  It is known to describe 
skyrmions \cite{Dadda}.  We call them spin-skyrmions since they are associated 
with the spin coherence.  The simplest excitation is given by one skyrmion 
with scale $\kappa $ sitting at $\bbox{x}=0$, whose wave function is
\begin{equation}
\F^{\spin}_{\sky}[\bbox{x}] = \prod _{r}\bpmatrix z_{r}\\ \kappa /2 \epmatrix^{\spin} \Laugh .
\label{SimplSkyrm}
\end{equation}
The skyrmion is reduced to the vortex in the limit $\kappa \rightarrow 0$.  The scale $\kappa $ is to 
be fixed dynamically.  The normalized spin (\ref{SigmaField}) is calculated from 
this wave function as
\begin{equation}
s^{x}_{\sky} = \sqrt {1-(s^{z})^{2}}\cos \theta ,\quad 
s^{y}_{\sky} =-\sqrt {1-(s^{z})^{2}}\sin \theta ,\quad 
s^{z}_{\sky} = {r^{2}-(\ell _{B}\kappa )^{2}\over r^{2}+(\ell _{B}\kappa )^{2}}.
\label{SkyrmSpin}
\end{equation}
The spin flips at the skyrmion center, $\bbox{s}=(0,0,-1)$ at $r=0$, while the 
spin-up-polarized ground state is approached away from it, $\bbox{s}=(0,0,1)$ for 
$r\gg \kappa \ell _{B}$.  

The soliton equation (\ref{SolitEquat}) follows together with the 
topological charge density,
\begin{equation}
Q_{0}(\bbox{x}) =Q^{P}_{0}(\bbox{x}) \equiv  -{i\over 2\pi }\sum _{\alpha  }\varepsilon _{jk}\partial _{j}\bigl(n^{\alpha *}\partial _{k}n^{\alpha }\bigr).
\label{SkyrmCharg}
\end{equation}
The topological charge is shown \cite{Dadda} to be identical with the Pontryagin 
number whose current density is
\begin{equation}
Q_{\mu }^{P}(\bbox{x}) = {1\over 8\pi }\varepsilon _{abc}\varepsilon _{\mu \nu \lambda }s^{a}\partial ^{\nu }s^{b}\partial ^{\lambda }s^{c} .
\label{PontrNumbe}
\end{equation}
Approximate solutions of the soliton equation (\ref{SolitEquat}) are constructed in 
the two limits, the large skyrmion limit ($\kappa \gg 1$) and the small skyrmion limit 
($\kappa \ll 1$).  First, in the large limit we can solve (\ref{SolitEquat}) iteratively, 
where the first order term is
\begin{equation}
\varrho  _{\sky}(\bbox{x}) \simeq  -\nu Q^{P}_{0}(\bbox{x}) = -{\nu \over \pi } {(\kappa \ell _{B})^{2}\over [r^{2}+(\kappa \ell _{B})^{2}]^{2}} .
\label{SkyrmDensi}
\end{equation}
This agrees with the formula due to Sondhi et al. \cite{SkyrmQH}.  However, in the 
small limit the topological charge $Q^{P}_{0}(\bbox{x})$ is localized within the core.  
Indeed, we have $Q_{0}^{P}(\bbox{x})\rightarrow \delta (\bbox{x})$ as $\kappa \rightarrow 0$ in (\ref{SkyrmDensi}), with which the 
solution of the soliton equation (\ref{SolitEquat}) is approximated by the vortex 
configuration $\varrho  _{\vor}(\bbox{x})$ in (\ref{BetteAppro}).  This is what we have expected 
since the skyrmion wave function is reduced to the vortex wave function in the 
limit $\kappa \rightarrow 0$, where there is no distinction between the U(1) and SU(2) 
excitations.  Since  the vortex may be considered as a small skyrmion limit, 
we do not make a clear distinction between them in what follows.

\section{Bilayer QH Systems}

We proceed to analyze bilayer QH systems.  In this section we freeze 
the spin degree of freedom.  We are interested in the so-called Halperin 
$(m,m,m)$ phase \cite{HalperinC}, or the interlayer-coherent phase, in which an 
interlayer quantum coherence develops spontaneously \cite{EIcoher,EzaIQC}.  We 
denote the electron field at the layer $\alpha (=1,2)$ by $\psi _{\alpha }$.  We call the layer 
$\alpha =1$ the front layer and the layer $\alpha =2$ the back layer.  It is convenient to 
introduce the pseudospin SU(2) structure by considering a two-component 
electron field $\Psi (\bbox{x})$ made of $\psi _{1}$ and $\psi _{2}$.  We use the pseudospin index 
$a=1,2,3$ and define the pseudospin density and the normalized pseudospin 
field by (\ref{SpinDensi}) and (\ref{SigmaField}), respectively.  In the 
interlayer-coherent phase, the dressed CB field $\Phi (\bbox{x})$ is defined precisely as 
in the monolayer QH ferromagnet.  The kinetic Hamiltonian of composite bosons 
is formally identical to that of the monolayer QH ferromagnet, and given by 
(\ref{HamilBL}).  Consequently, the LLL condition is identical to (\ref{LLLcondiBL}), 
and the Hilbert space made of the states in the \LLL\ is identical to that of 
the monolayer QH ferromagnet.  We make a full use of this mathematical 
identity \cite{EzaIQC} to investigate the interlayer-coherent phase.

The Coulomb interaction is summarized into the two terms \cite{EzaIQC},
\begin{align}
H_{C}^{+}&= {1\over 2}\int \dx\dx V^{+}(\bbox{x}-\bbox{y})\varrho  (\bbox{x})\varrho  (\bbox{y}), \label{HamilPlus}\\
H_{C}^{-}&= 2\int \dx\dx V^{-}(\bbox{x}-\bbox{y})\Delta S^{3}(\bbox{x})\Delta S^{3}(\bbox{y}), \label{HamilCapac}
\end{align}
where $V^{\pm }(\bbox{x})=(e^{2}/2\varepsilon )\bigl(|\bbox{x}|^{-1} \pm  (|\bbox{x}|^{2}+d^{2})^{-1/2}\bigr)$ with the interlayer separation 
$d$: $\varrho  (\bbox{x})$ is the deviation of the total electron density.  The Coulomb 
energy $H_{C}^{+}$ is the driving force to realize the QH system.  The term $H_{C}^{-}$ 
describes the capacitive charging energy between the two layers.  The 
deviation of the pseudospin density from the ground-state value is
\begin{equation}
\Delta S^{a}(\bbox{x}) \equiv  {1\over 2}\bigl[\rho (\bbox{x})s^{a}(\bbox{x}) - \rho _{0}s^{a}_{0}\bigr] ,
\label{PspinDevia}
\end{equation}
where $s^{a}(\bbox{x})$ is the normalized pseudospin and $s_{0}^{a}$ is its average value in 
the ground state: See (\ref{GrounPspin}).  On the other hand, the tunneling term is 
given by
\begin{equation}
H_{T} = - \DSAS \int \dx \Delta S^{1}(\bbox{x}) .
\label{TunneA}
\end{equation}
The tunneling term produces an energy gap $\DSAS$ between the symmetric and 
antisymmetric states.  The pseudospin (interlayer) coherence develops provided 
the capacitance term $H_{C}^{-}$ and the tunneling term $H_{T}$ are reasonably small. 

There is one additional degree of freedom in the bilayer system.  By 
applying bias voltages to the two layers, we can freely control the electron 
density $\rho _{0}^{\alpha }$ in each layer \cite{SawaPRLa}.  Namely, the direction of the 
pseudospin polarization is controllable \cite{EzaIQC}.  Accordingly, the ground 
state wave function is given by
\begin{equation}
\F^{\ppin}_{\text{g}}[\bbox{x}] = \prod _{r}\bpmatrix \sqrt {1+\sigma _{0}}\\ \sqrt {1-\sigma _{0}}\epmatrix_{r}^{\ppin} \Laugh ,
\label{HalpeMM}
\end{equation}
where the pseudospinor is common to all electrons and the parameter $\sigma _{0}$ is a 
real constant ($|\sigma _{0}|\leq 1$); the index $\textit{ppin}$ denotes the pseudospin.  
Note that the ground state (\ref{HalpeMM}) has been chosen so as to minimize the 
tunneling energy.  The density ratio between the two layers is
\begin{equation}
{\rho ^{1}_{0}\over \rho ^{2}_{0}} = {1+\sigma _{0}\over 1-\sigma _{0}} .
\label{HalpeDensi}
\end{equation}
The normalized pseudospin is calculated from the wave function (\ref{HalpeMM}) as
\begin{equation}
s^{1}_{0}(\bbox{x})= \sqrt {1-\sigma _{0}^{2}} ,\quad \quad  s^{2}_{0}(\bbox{x})= 0,\quad \quad  s^{3}_{0}(\bbox{x})= \sigma _{0} .
\label{GrounPspin}
\end{equation}
The balanced configuration is realized when $\sigma _{0}=0$, where the symmetric state 
is the ground state and described by (\ref{HalpeMM}) with $\sigma _{0}=0$.

We analyze charged excitations on the ground state (\ref{HalpeMM}).  The 
easiest way is to use a mapping between the bilayer pseudospin state and the 
monolayer spin state \cite{EzaIQC}.  The mapping is established by the matrix,
\begin{equation}
T_{\sigma } = {1\over \sqrt {2}}
\begin{pmatrix}\sqrt {1+\sigma _{0}} & \sqrt {1-\sigma _{0}}\\
               \sqrt {1-\sigma _{0}} &-\sqrt {1+\sigma _{0}}\end{pmatrix} .
\end{equation}
It is easy to see that $T_{\sigma }\F_{\text{g}}^{\ppin}=\F^{\spin}_{\text{g}}$, where 
$\F^{\ppin}_{\text{g}}$ is the ground state (\ref{HalpeMM}) in the unbalanced bilayer 
system and $\F^{\spin}_{\textit{g}}$ is the ground state (\ref{GrounWave}) in the 
monolayer QH ferromagnet.  All excitations are mapped by this transformation 
between the two systems.  Consequently, the skyrmion excitation on the ground 
state (\ref{HalpeMM}) is obtained from that on (\ref{GrounWave}) as 
$\F^{\ppin}_{\sky}=T_{\sigma }^{\dagger }\F^{\spin}_{\sky}$ with (\ref{SimplSkyrm}), or
\begin{equation}
\F^{\ppin}_{\sky} = \prod _{r}\bpmatrix z_{r}\sqrt {1+\sigma _{0}}+(\kappa /2)\sqrt {1-\sigma _{0}} \\ 
                      z_{r}\sqrt {1-\sigma _{0}}-(\kappa /2)\sqrt {1+\sigma _{0}}\epmatrix^{\ppin} \Laugh.
\label{WaveSkyrmBi}
\end{equation}
This is the pseudospin-skyrmion associated with the pseudospin coherence.  The 
normalized pseudospin (\ref{SigmaField}) is easily calculable from this wave 
function with the aid of (\ref{GenerSkyrm}).  It turns out that one skyrmion 
excitation consists of two parts, one on the front layer and the other on the 
back layer.  

\section{Bilayer QH Ferromagnets}

We finally analyze the bilayer QH system with spins included.  The 
lowest Landau level contains four energy levels corresponding to the two 
layers and the two spin states.  In the balanced configuration the ground 
state is the symmetric and spin-up state, which is separated by the Zeeman gap 
$g^{*}\mu _{B}B$ and/or the tunneling gap $\DSAS$ from the one-particle excited states.  
In the unbalanced configuration at $\nu =1/m$ the QH state is given by
\begin{equation}
\F[\bbox{x}] = \prod _{r}\bpmatrix 1\\ 0\epmatrix^{\spin}_{r}
\bpmatrix \sqrt {1+\sigma _{0}}\\ \sqrt {1-\sigma _{0}}\epmatrix^{\ppin}_{r} \Laugh ,
\label{QHferroBL}
\end{equation}
as a tensor product of the spinor and the pseudospinor.  When both of the 
interactions are much smaller than the Coulomb energy, two kinds of quantum 
coherences develop spontaneously upon this ground state.  

We consider the filling factor $\nu =1$ and $\nu =2$.  There are two cases.  
(A) When the Zeeman gap is smaller than the tunneling gap, spin-skyrmions are 
excited at $\nu =1$ while pseudospin-skyrmions are excited at $\nu =2$.  (B) When 
the Zeeman gap is larger than the tunneling gap, pseudospin-skyrmions are 
excited at $\nu =1$ while spin-skyrmions are excited at $\nu =2$.  

We are interested in the activation energy $\Delta _{\text{act}}(\sigma _{0})$ as a 
function of the density imbalance parameter $\sigma _{0}$.  Though $\Delta _{\text{act}}(\sigma _{0})$ is 
sensitive to impurities in samples, the difference $\Delta _{\text{act}}(\sigma _{0})-
\Delta _{\text{act}}(0)$ is not so.  The excitation energy of the spin-skyrmion depends 
on $\sigma _{0}$ only through the capacitance term (\ref{HamilCapac}), but that of the 
pseudospin-skyrmion depends also on the tunneling term (\ref{TunneA}).  The 
capacitive charging energy arises since one skyrmion consists of two charge 
excitations, one in the front layer ($q_{e}^{1}$) and the other on the back layer 
($q_{e}^{2}$).  It carries the total electric charge (\ref{TotalCharg}) with 
$q_{e}=q_{e}^{1}+q_{e}^{2}=\nu e$.  

We analyze spin-skyrmions.  Though the normalized pseudospin 
(\ref{GrounPspin}) is not affected by the spin-skyrmion excitation, it induces a 
deviation of the pseudospin (\ref{PspinDevia}),
\begin{equation}
\langle \Delta S^{3}(\bbox{x})\rangle  \equiv  {1\over 2}\langle \varrho  _{\sky}(\bbox{x})\rangle s^{3}_{0} = {\sigma _{0}\over 2}\varrho  _{\sky}(\bbox{x}) .
\end{equation}
The charge difference is
\begin{equation}
q_{e}^{1}-q_{e}^{2} = -2e \int \dx \langle \Delta S^{3}(\bbox{x})\rangle  = \nu \sigma _{0}e ,
\end{equation}
independently of the details of excitations.  Hence, the charge on the front 
layer is $q_{e}^{1}={1\over 2}\nu e(1+\sigma _{0})$, while that on the back layer is $q_{e}^{2}={1\over 2}\nu e(1-
\sigma _{0})$.  The ratio $q_{e}^{1}/q_{e}^{2}$ is the same as the ratio (\ref{HalpeDensi}) of the 
electron densities on the two layers, as is expected.  The capacitive charging 
energy (\ref{HamilCapac}) is estimated as
\begin{equation}
\langle H_{C}^{-}\rangle  = \alpha _{C} {e^{2}\sigma _{0}^{2}\over \varepsilon \ell _{B}} ,
\label{ActivCapac}
\end{equation}
with a numerical constant $\alpha _{C}$ at a fixed value of $d/\ell _{B}$ (the ratio of the 
interlayer distance $d$ and the magnetic length $\ell _{B}$).  Its value depends on 
the details of the excitation.  The pseudospin-skyrmion is similarly 
analyzed, and the same formula as (\ref{ActivCapac}) is obtained for its charging 
energy.

\begin{figure}
\def\psbox#1#2{\includegraphics*[width=#1]{#2}}
\psbox{120mm}{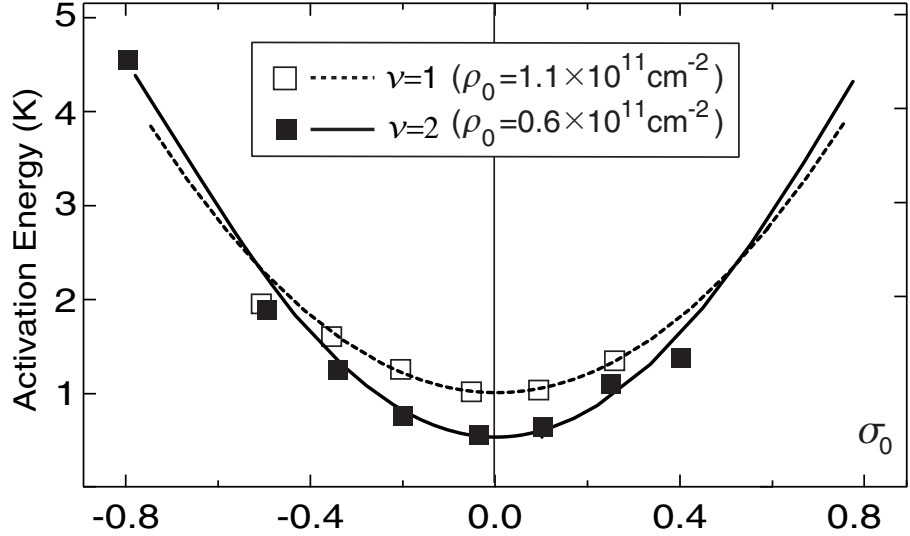}\label{ExcEneBLPS}\caption{%
The excitation energy (in K) of one skyrmion is calculated as a function of 
the imbalance parameter $\sigma _{0}$, and compared with the experimental data taken 
from Ref.\cite{SawaPRLa}.  We have adjusted the value at the balanced point 
($\sigma _{0}=0$) to the experimental one.  
}
\end{figure}

We have compared the above results with the experimental data taken 
from Ref.\cite{SawaPRLa}, where $g^{*}\mu _{B}B/\DSAS\simeq 1/4$ at $B=5$ Tesla.  See 
Fig.\ref{ExcEneBLPS}.  At $\nu =1$ we have fitted the activation-energy data by 
assuming a spin-skyrmion excitation with size $\kappa =2$, where the relevant 
parameters are $d=231\AA$ and $\ell _{B}=120\AA$.  At $\nu =2$ we have fitted the data 
by assuming a pseudospin-skyrmion excitation with size $\kappa =1.65$ and charge 
$2e$, where the relevant parameters are $d=231\AA$, $\ell _{B}=228\AA$ and 
$\DSAS=6$K.  The agreements are quite good, as would imply that the present 
picture of the bilayer QH ferromagnet is basically correct.  Detailed analysis 
will be published in a separate paper \cite{ESSa}.

\section{Discussions}

We have studied both the monolayer and bilayer QH ferromagnets based on 
the improved CB theory.  We have investigated soliton excitations confined 
to the \LLL.  The semiclassical approximation is powerful when the CB wave 
function $\F_{\varphi }$ is factorizable.  This allows us to analyze quasiholes 
successfully.  It is quite difficult to make a similar analysis of 
quasielectrons, for which $\F_{\varphi }$ is not factorizable.  In comparing our 
theoretical results with experimental data, we have made a physical assumption 
that the excitation energy of one quasielectron is approximately the same as 
that of one quasihole.  In order to discuss quasielectrons, it would be 
necessary to make the LLL projection after exciting them.  We hope to discuss 
on this point in a future paper.  

I would like to thank K. Sasaki and A. Sawada for various discussions 
on the subject.  A partial support is acknowledged from a Grant-in-Aid for the 
Scientific Research from the Ministry of Education, Science, Sports and 
Culture.

\vspace{20mm}


\newpage
\begin{thebibliography}{99}
\bibitem{FQHEbook}
{\it The Quantum Hall Effect}, edited by S. Girvin and 
R. Prange (Springer-Verlag, New York, 1990), 2nd ed.
\bibitem{LGCSx}
S.M. Girvin and A.H. MacDonald, Phys.\ Rev.\ Lett.\  {\bf 58}, (1987) 1252;
S.C. Zhang, T.H. Hansen and S. Kivelson, Phys.\ Rev.\ Lett.\  {\bf 62}, (1989) 82;
Z.F. Ezawa and A. Iwazaki, Phys.\ Rev.\ B \  {\bf 43}, (1991) 2637;
Z.F. Ezawa, M. Hotta and A. Iwazaki, Phys.\ Rev.\ B \  {\bf 46}, (1992) 7765;
S.C. Zhang, Int.\ J.\ Mod.\ Phys.\  B {\bf 6}, (1992) 25.
\bibitem{GirvinAA}
S.M. Girvin, in Ref.[1], Chap 10.
\bibitem{ReadA}
N. Read, Phys.\ Rev.\ Lett.\  {\bf 62}, (1989) 86.
\bibitem{RajaramanCB}
R. Rajaraman and S.L. Sondhi, Int.\ J.\ Mod.\ Phys.\  B {\bf 10}, (1996) 793.
\bibitem{JainCF}
J.K. Jain, Phys.\ Rev.\ Lett.\  {\bf 63}, (1989) 199.
\bibitem{OtherCF}
A. Lopez and E. Fradkin, Phys.\ Rev.\ B \  {\bf 44}, (1991) 5246;
V. Kalmeyer and S.C. Zhang, Phys.\ Rev.\ B \  {\bf 46}, (1992) 9889;
B.I. Halperin, P.A. Lee and N. Read, Phys.\ Rev.\ B \ {\bf 47}, (1993) 7312.
\bibitem{EzaICBa}
Z.F. Ezawa, Physica B {\bf 249--251}, (1998) 841
\bibitem{NewApproCP}
R. Shankar and G. Murthy, Phys.\ Rev.\ Lett.\  {\bf 79}, (1997) 4437;
D.H. Lee, cond-mat/9709233;
V. Pasquier and F.D.M. Haldane, cond-mat/9712169.
\bibitem{SkyrmQH}
S.L. Sondhi, A. Karlhede, S. Kivelson and E.H. Rezayi, 
Phys.\ Rev.\ B \  {\bf 47}, (1993) 16419;
D.H. Lee and C.L. Kane, Phys.\ Rev.\ Lett.\  {\bf 64}, (1990) 1313.
\bibitem{SkyrmExper}
S.E. Barrett, G. Dabbagh, L.N. Pfeiffer, K.W. West and R. Tycko,
Phys.\ Rev.\ Lett.\  {\bf 74}, (1995) 5112; 
A. Schmeller, J.P. Eisenstein, L.N. Pfeiffer and K.W. West,
Phys.\ Rev.\ Lett.\  {\bf 75}, (1995) 4290;
E.H. Aifer, B.B. Goldberg and D.A. Broido,
Phys.\ Rev.\ Lett.\  {\bf 76}, (1996) 680;
D.K. Maude, M. Potemski, J.C. Portal, M. Henini, L. Eaves,
G. Hill and M.A. Pate,
Phys.\ Rev.\ Lett.\  {\bf 76}, (1996) 4604.
\bibitem{EIcoher}
Z.F. Ezawa and A. Iwazaki, Int.\ J.\ Mod.\ Phys.\  B {\bf 6}, (1992) 3205;
Phys.\ Rev.\ B \  {\bf 47}, (1993) 7295; 
Phys.\ Rev.\ B \  {\bf 48}, (1993) 15189; 
\bibitem{EzaIQC}
Z.F. Ezawa, Phys.\ Lett.\  A {\bf 229}, (1997) 392;
Z.F. Ezawa, Phys.\ Rev.\ B \  {\bf 55}, (1997) 7771.
\bibitem{MG}
K. Moon, H. Mori, K. Yang, S.M. Girvin, A.H. MacDonald, 
L. Zheng, D. Yoshioka and S.C. Zhang, Phys.\ Rev.\ B \  {\bf 51}, (1995) 5138.
\bibitem{Sheena}
S.Q. Murphy, J.P. Eisenstein, G.S. Boebinger, L.N. Pfeiffer 
and K.W. West, Phys.\ Rev.\ Lett.\  {\bf 72}, (1994) 728.
\bibitem{SawaPRLa}
A. Sawada, Z.F. Ezawa, H. Ohno, Y. Horikoshi, Y. Ohno,
S. Kishimoto, F. Matsukura, M. Yasumoto, and A. Urayama,
Phys.\ Rev.\ Lett.\  {\bf 80}, (1998) 4534.
\bibitem{LaughlinA}
R.B. Laughlin, Phys.\ Rev.\ Lett.\  {\bf 50}, (1983) 1395.
\bibitem{Dadda}
A. D'Adda, A. Luscher and P. DiVecchia, Nucl.\ Phys.\  {\bf B146}, (1978) 63.
\bibitem{AnyonExperA}
L. Saminadayar, D.C. Glattli, Y. Jin and B. Etienne,
Phys.\ Rev.\ Lett.\  {\bf 79}, (1997) 2526;
R. de-Picciotto, M. Reznikov, M. Heiblum, V. Umansky, 
G. Bunin andd D. Mahalu,
Nature {\bf 386}, (1997) 162.
\bibitem{HalperinC}
B.I. Halperin, Helv. Phys. Acta {\bf 56}, (1983) 75.
\bibitem{ESSa}
Z.F. Ezawa, K. Sasaki and A. Sawada, in preparation.
\end{thebibliography}
\end{document}